\documentclass[12pt]{iopart}
\usepackage{iopams} 
\usepackage{graphicx}
\usepackage{amssymb}
\usepackage{amsfonts}
\newcommand{\be}{\begin{equation}}
\newcommand{\ee}{\end{equation}}
\newcommand{\bd}{\begin{displaymath}}
\newcommand{\ed}{\end{displaymath}}
\newcommand{\ba}{\begin{eqnarray}}
\newcommand{\ea}{\end{eqnarray}}
\begin{document}


\font\mybig=cmbx12 at 14pt
\def\half{{\textstyle{1\over2}}}
\def\halfs{{\scriptstyle{1\over2}}}

\font\mybf=cmbx12 at 14pt
\font\mymi=cmmi12 at 14pt
\font\myssl=cmss12 at 14pt
\font\myrm=cmr10
\font\myrms=cmr7
\font\mybfs=cmbx7
\font\myss=cmss10
\font\mysss=cmss10 at 7pt
\font\mybs=cmbx10
\font\mybss=cmbx7
\font\mymib=cmmib10
\font\mymibs=cmmib10 at 7pt
\font\bfrak=eufb10
\def\mss#1{\hbox{\myss#1}}
\def\msss#1{\hbox{\mysss#1}}
\def\bGamma{\hbox{\mybs\char'00}}
\def\bsGamma{\hbox{\mybss\char'00}}
\def\bsigma{\hbox{\mymib\char'33}}
\def\bssigma{\hbox{\mymibs\char'33}}
\def\bsigma{\hbox{\mymib\char'33}}
\def\btau{\hbox{\mymib\char'34}}
\def\bmu{\hbox{\mymib\char'26}}
\def\bnu{\hbox{\mymib\char'27}}
\def\one2{\pmatrix{1&0\cr0&1\cr}}
\def\stt{\vtop to 0.1in{}}
\catcode`\@=11
\def\setboxz@h{\setbox\z@\hbox}
\def\wdz@{\wd\z@}
\def\boxz@{\box\z@}
\def\setbox@ne{\setbox\@ne}
\def\wd@ne{\wd\@ne}
\def\binrel@#1{\setboxz@h{\thinmuskip0mu
  \medmuskip\m@ne mu\thickmuskip\@ne mu$#1\m@th$}%
 \setbox@ne\hbox{\thinmuskip0mu\medmuskip\m@ne mu\thickmuskip
  \@ne mu${}#1{}\m@th$}%
 \setbox\tw@\hbox{\hskip\wd@ne\hskip-\wdz@}}
\def\overset#1#2{\binrel@{#2}\ifdim\wd\tw@<\z@
 \mathbin{\mathop{\kern\z@#2}\limits^{#1}}\else\ifdim\wd\tw@>\z@
 \mathrel{\mathop{\kern\z@#2}\limits^{#1}}\else
 {\mathop{\kern\z@#2}\limits^{#1}}{}\fi\fi}
\def\underset#1#2{\binrel@{#2}\ifdim\wd\tw@<\z@
 \mathbin{\mathop{\kern\z@#2}\limits_{#1}}\else\ifdim\wd\tw@>\z@
 \mathrel{\mathop{\kern\z@#2}\limits_{#1}}\else
 {\mathop{\kern\z@#2}\limits_{#1}}{}\fi\fi}
\catcode`\@=12

\font\mathaa=msam10
\font\mathaas=msam10 at 7pt
\def\leqslant{\,\hbox{\mathaa\char"36}\,}
\def\leqsls{\,\hbox{\mathaas\char"36}\,}
\def\geqslant{\,\hbox{\mathaa\char"3E}\,}
\def\geqsls{\,\hbox{\mathaas\char"3E}\,}
\def\frac#1#2{{#1\over#2}}
\def\sbbR{{\mathcal S}(\mathbb{R})}
\def\wphi{\widehat{\phi}}
\def\wpsi{\widehat{\psi}}
\def\bT{{\boldsymbol{T}}}
\def\bt{{\boldsymbol{t}}}
\def\vp{^{\vphantom{y}}}
\def\P{{\rm P}}

\def\bx{{\bf x}}
\def\bX{{\bf X}}
\def\bY{{\bf Y}}
\def\bZ{{\bf Z}}
\def\bpsi{\hbox{\mymib\char'40}}
\def\xy{{\textstyle{\frac{x}{y}}}}
\def\XY{{\displaystyle{\frac{x}{y}}}}
\def\q{{q^{-1}}}
\def\wb{{\overline W}}
\def\bone{{\bf 1}}
\def\bzero{{\bf 0}}

\def\boxit#1{\hbox{\lower3pt\vbox{\hrule\hbox{\vrule\kern3pt
  \vbox{\kern3pt\hbox{#1}\kern3pt}\kern3pt\vrule}\hrule}}}
\def\bigboxit#1#2{\hbox{\lower#1pt\vbox{\hrule\hbox{\vrule\kern3pt
  \vbox{\kern3pt\hbox{#2}\kern3pt}\kern3pt\vrule}\hrule}}}


\title[Odd-Even Root-of-Unity Problem and Chiral Potts Model]%
{Integrable Chiral Potts Model and the Odd-Even Problem in Quantum Groups
at Roots of Unity}

\author{Helen Au-Yang and Jacques H.H. Perk}
\address{Department of Physics, Oklahoma State University, 
145 Physical Sciences, Stillwater, OK 74078-3072, USA}
\ead{helenperk@yahoo.com, perk@okstate.edu}

\begin{abstract}
At roots of unity the $N$-state integrable chiral Potts model and the six-vertex model descend from each other with the $\tau_2$ model as the intermediate. We shall discuss how different gauge choices in the six-vertex model lead to two different quantum group constructions with different $q$-Pochhammer symbols, one construction only working well for $N$ odd, the other equally well for all $N$. We also address the generalization based on the sl$(m,n)$ vertex model.
\end{abstract}
\maketitle


\section {Introduction}

Ever since the discovery \cite{AMPTY} of the Yang--Baxter integrable
chiral Potts model in 1986 with spectral variables (rapidities) living on
higher-genus curves, many papers have been written to understand it better,
including its first complete explicit parametrization \cite{BPA}.%
\footnote{The early history has been reviewed recently in \cite{Perk}.}\
It became soon clear that the model has to be related to the six-vertex
model by some cyclic, rather than highest/lowest-weight, representation. Such a quantum-group structure in mathematics has been advocated by
de Concini and Kac \cite{dCK} around 1990. They worked out the case of
primitive $\ell$-th roots-of-one with $\ell$ odd. The case $\ell$ even
was left as an open problem by them.

For the chiral Potts model the quantum-group construction was first
worked out by Bazhanov and Stroganov \cite{BS} for the number of states
per spin $N$ being odd, starting from the six-vertex model. Here $N$ is
the $\ell$ of \cite{dCK}. As there is no clear distiction between odd
and even $N$ in \cite{BPA}, a different construction was given valid
for all $N$ starting from chiral Potts \cite{BBP}. The difference between
these two $U_q(\widehat{\mathfrak{sl}}(2))$ constructions has been
discussed recently in section 3 of \cite{Perk} and section 1.3 of
\cite{APCSOS}. As \cite{BBP} is more difficult to read, many authors
prefer to use the \cite{BPA} approach and are consequently limited
to the $N$ odd case, see e.g.\ \cite{MNP} and references cited. It may,
therefore, be useful to compare the two approaches in more detail.
In doing so, we shall compare the approaches of \cite{BS} and \cite{BBP}
and also compare with Korepanov's derivation \cite{Korepanov,Korepanov2}
of his version of the $\tau_2$ model.

We shall also address the two constructions of the $U_q(\widehat{\mathfrak{sl}}(n))$ generalization of the chiral Potts model,
which can be seen as a special $n-1$ layer $N$-state chiral Potts model.
The derivation in \cite{DJMM} depends on $N$ being odd, whereas \cite{BKMS}
is valid for all $N$.

The quantum group structure has become important in our later works, as
it leads to a simpler proof of the needed quantum Serre relations,
needed for example in proofs of conjectures on free parafermions in the
$\tau_2$ model \cite{Fendley,Baxter,AYP,AYP2}.


\section{Constructions based on sl(m,n) vertex model}

\begin{figure}[htb]
\begin{center}
     \includegraphics[width=0.15\hsize]{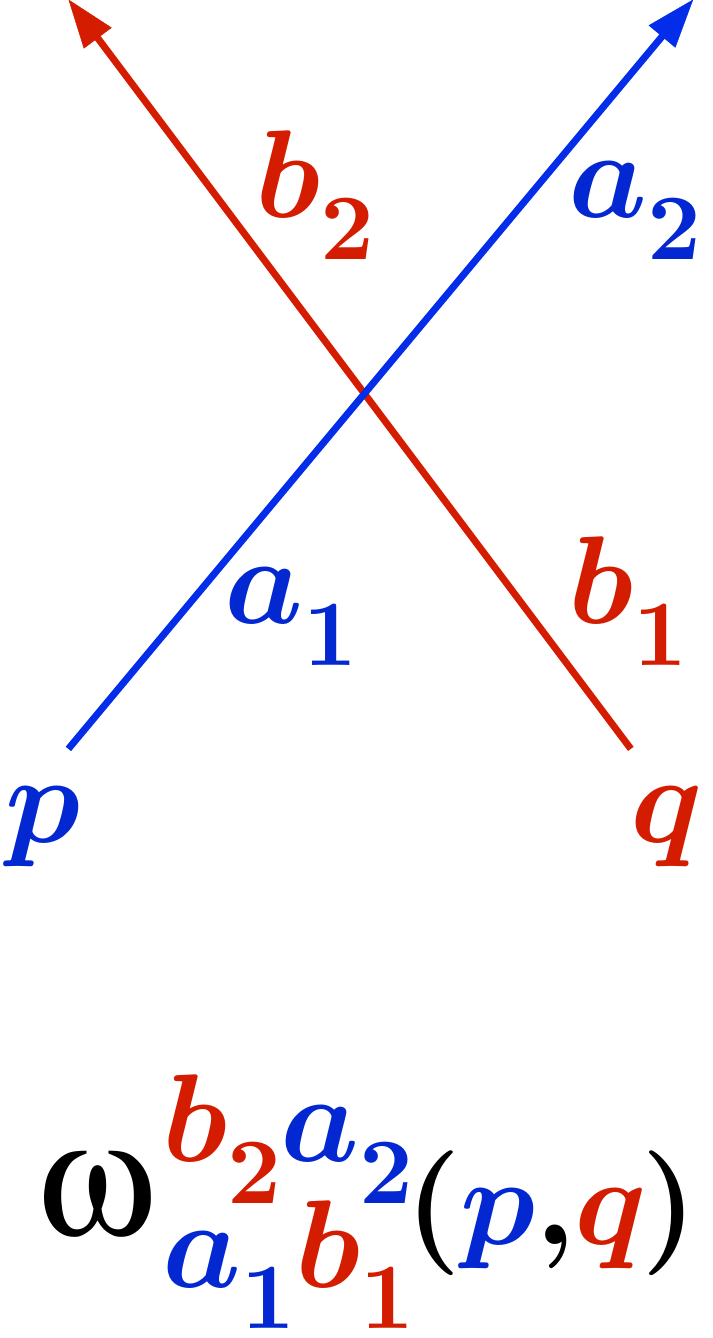}
\hspace*{2.5cm}
        \includegraphics[width=0.55\hsize]{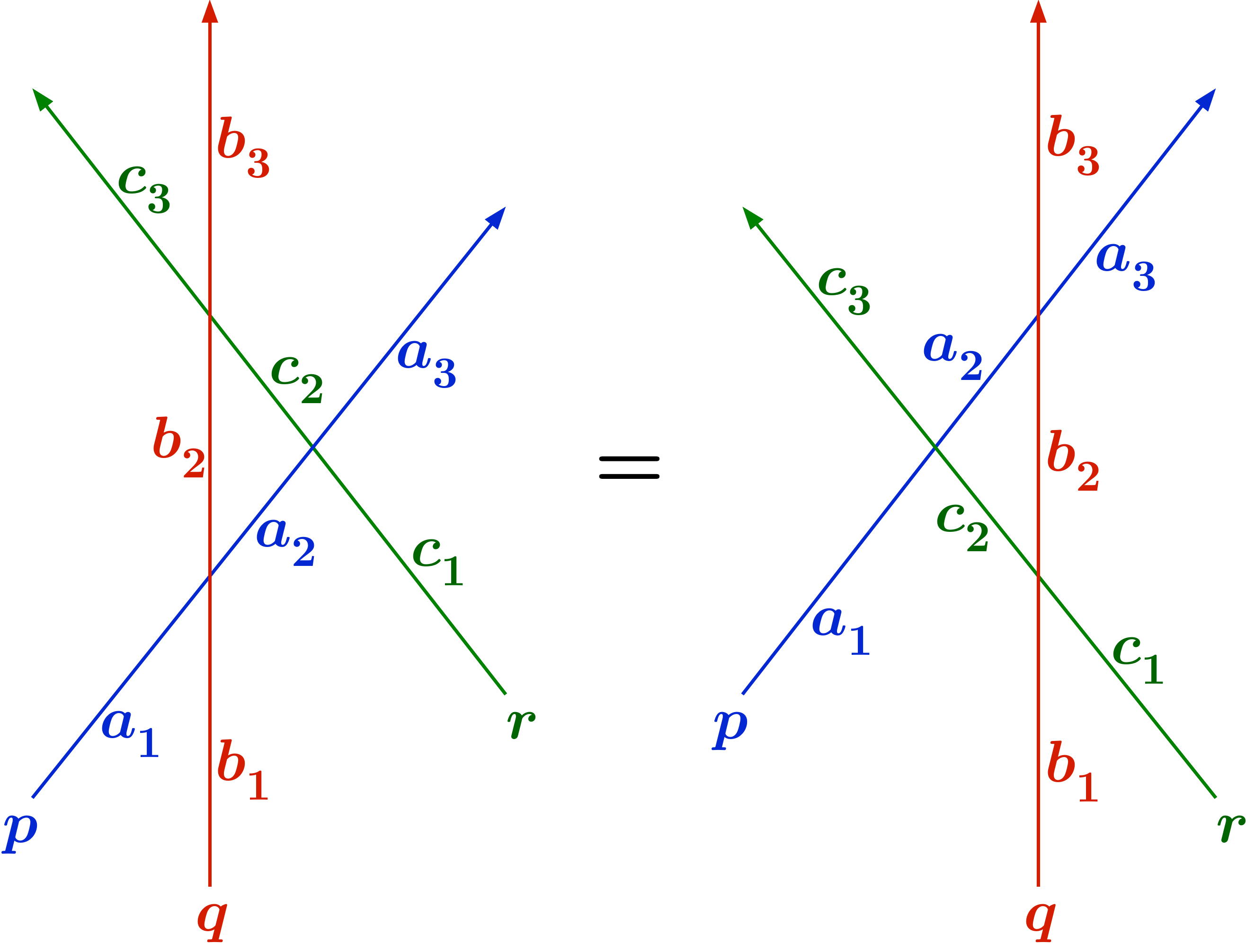}
\end{center}
\caption{(Color online) Vertex-model Boltzmann weights (R-matrix)
in graphical representation on the left and corresponding Yang--Baxter
equation on the right. Here $a_j,b_j,c_j$ are state variables living
on the edges and $p,q,r$ are line variables (rapidities). The
states $a_2,b_2,c_2$ are summed over.}
\label{fig1}
\end{figure}

In order to construct chiral Potts models based on quantum group
$U_q(\widehat{\mathfrak{sl}}(m,n))$, we start with the fundamental
R-matrix given through the ${\rm sl}(m,n)$ vertex model of \cite{PS}.
${\rm sl}(m,n)$ vertex model. This R-matrix, solving the Yang--Baxter
equation in Fig.~\ref{fig1}, is best given in the parametrization of
\cite{APYB}, with the non-zero weights being
\ba
\fl\bomega^{aa}_{aa}(p,q)=
\mathcal{N}\sinh\big(\eta+\varepsilon_a(p_0-q_0)\big)\,\displaystyle\frac{p_{+a}q_{-a}}{q_{+a}p_{-a}},
\qquad(a=1,\cdots,m+n);
\label{eq1a}\\
\fl\bomega^{ab}_{ba}(p,q)=\mathcal{N}\,G_{ab}\sinh\big(p_0-q_0\big)\,
\displaystyle\frac{p_{+a}q_{-b}}{q_{+b}p_{-a}},
\qquad(a\ne b,\quad a,b=1,\cdots,m+n);
\label{eq1b}\\
\fl\bomega^{ba}_{ba}(p,q)=
\mathcal{N}\,{\rm e}^{(p_0-q_0){\rm sign}(a-b)}\sinh\big(\eta\big)\,
\displaystyle\frac{p_{+b}q_{-a}}{q_{+b}p_{-a}},
\qquad(a\ne b,\quad a,b=1,\cdots,m+n).
\label{eq1c}
\ea
Here we have $(2m+2n+1)$-component rapidities $p$ and $q$, with
$p_{\pm i}$ and $q_{\pm i}$ for $i\ne0$ being gauge parameters.
Also we have $m$ plus signs and $n$ minus signs, which we can order
$\varepsilon_a=+1$ ($a=1,\cdots,m$),
$\varepsilon_a=-1$ ($a=m+1,\cdots,m+n$). Furthermore, $\mathcal{N}$
is an arbitrary normalization, $\eta$ is a constant and the $G_{ab}$
are constant twist parameters satisfying $G_{ab}G_{ba}=1$.%
\footnote{We can make $G_{ab}\equiv1$ by suitable changes of
the gauge rapidities only when $m+n=2$.}

\begin{figure}[htb]
\begin{center}
  \vspace*{0pt}
     \includegraphics[width=0.5\hsize]{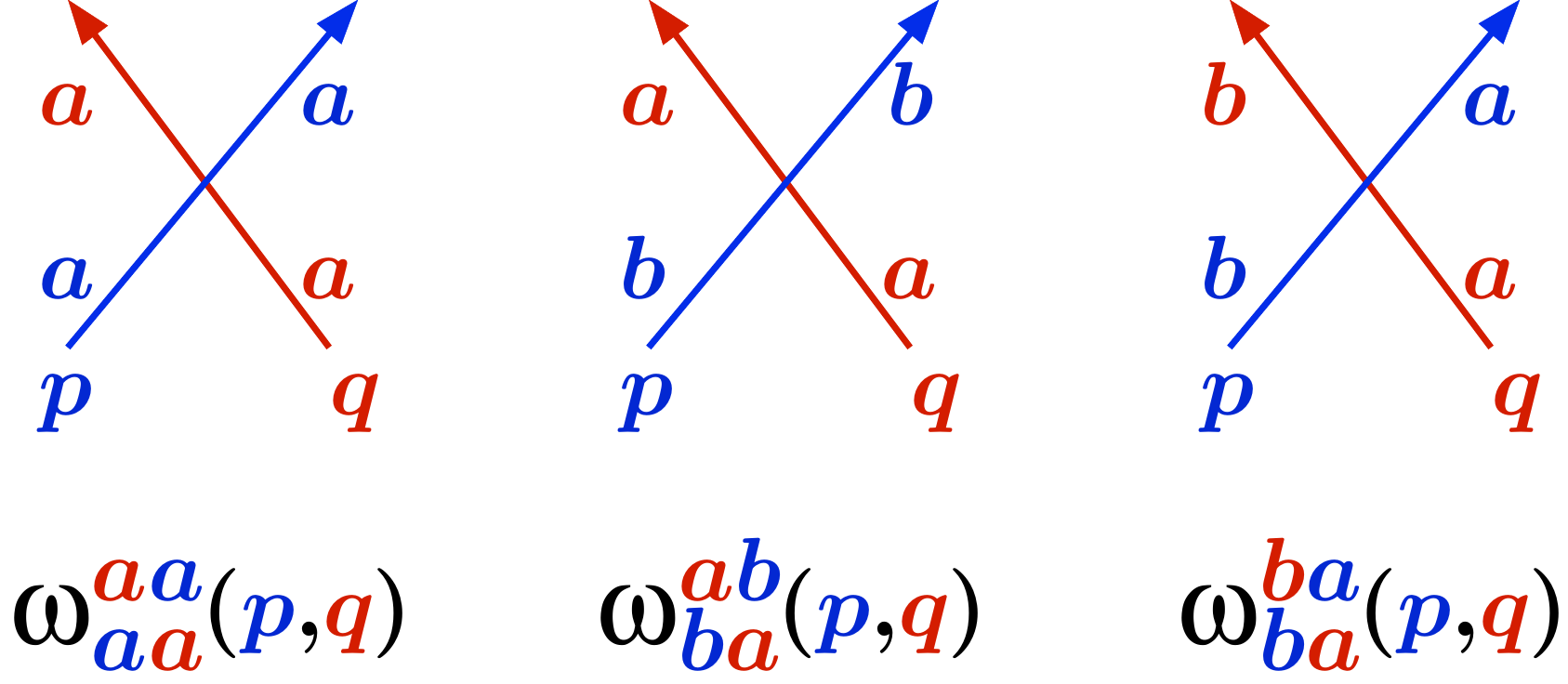}
\end{center}
\caption{(Color online) The three types of non-zero Boltzmann weights
of the sl$(m,n)$ vertex model of \cite{PS}.}
\label{fig2}
\end{figure}

\noindent We change the variables according to
\be
q\equiv\mathrm{e}^{2\eta},\quad
x\equiv\mathrm{e}^{2q_0},\quad y\equiv\mathrm{e}^{2p_0},\quad
\mathcal{N}\,{q^{1/2}\over\raise1pt\hbox{$2$}}
\Big(\frac{y}{x}\Big)^{\!1/2}\!\equiv\mathcal{N}',
\ee
in order to change the additive rapidities $p_0$ and $q_0$ to
multiplicative rapidities $x$ and $y$. Thus we get
\ba
\bomega^{aa}_{aa}(p,q)=\cases{
\mathcal{N}'\Big(1-q^{-1}\XY\Big)\displaystyle\frac{p_{+a}q_{-a}}{q_{+a}p_{-a}},&if $\varepsilon_a=+1$,\cr
\mathcal{N}'\Big(\XY-q^{-1}\Big)\displaystyle\frac{p_{+a}q_{-a}}{q_{+a}p_{-a}},&if $\varepsilon_a=-1$,}\\
\bomega^{ab}_{ba}(p,q)=\mathcal{N}'G_{ab}\,q^{-1/2}\Big(1-\XY\Big)\displaystyle\frac{p_{+a}q_{-a}}{q_{+a}p_{-a}},\\
\bomega^{ba}_{ba}(p,q)=\cases{\mathcal{N}'(1-q^{-1})\XY\displaystyle\frac{p_{+b}q_{-a}}{q_{+b}p_{-a}},&if $a<b$,\cr
\mathcal{N}'(1-q^{-1})\displaystyle\frac{p_{+b}q_{-a}}{q_{+b}p_{-a}},&if $a>b$.}
\ea
We reduce this to the root-of-unity case, if we set
$\eta=j\pi\mathrm{i}/N$, or $q=\mathrm{e}^{2j\pi\mathrm{i}/N}$. When
$j$ and $N$ are relative prime, $q$ is a primitive root of one. One
can then proceed to cyclic representations of the quantum group
$U_q(\widehat{\mathfrak{sl}}(n))$, provided one deals with the
integer and half-integer powers of $q$ that may occur. The approach
in \cite{DJMM} restricted $N$ to be odd, so that $q^{1/2}=-q^{(N+1)/2}$
and one only has integer powers of $q$ to deal with.

If $N\geqslant3$, there is no choice of $p_{\pm a}$, $q_{\pm a}$ and
$\mathcal{N}'$ that can eliminate the half-integer powers of $q$. So,
let us set $p_{\pm a}=q_{\pm a}\equiv1, (a\ne0),$ and $\mathcal{N}'\equiv1$.
Then we arrive at
\ba
\bomega^{aa}_{aa}(p,q)=\cases{
1-q^{-1}\XY,&if $\varepsilon_a=+1$,\cr
\XY-q^{-1},&if $\varepsilon_a=-1$,}\\
\bomega^{ab}_{ba}(p,q)=G_{ab}\,q^{-1/2}\Big(1-\XY\Big)=
\cases{1-\XY,&if $a>b$,\cr q^{-1}\Big(1-\XY\Big),&if $a<b$,}
\label{Gab}\\
\bomega^{ba}_{ba}(p,q)=\cases{(1-q^{-1})\XY,&if $a<b$,\cr
1-q^{-1},&if $a>b$,}
\ea
provided we also choose $G_{ab}=q^{\pm{\rm sign}(a-b)/2}$,
($G_{ab}G_{ba}=1$), in (\ref{Gab}). Then any $\bomega_{ab}^{cd}(p,q)$
is a linear combination of $1,q^{-1},\xy,q^{-1}\xy$ only! This is how
\cite{BKMS} overcame the odd-even $N$ problem, albeit that they have not
spelled this out so explicitly.

Just choosing a more asymmetric R-matrix, or equivalently a different
coproduct, one can treat the even and odd $N$ cases in a uniform way.
This was also noted in \cite{Perk} for the $n=2$ case, with the
fundamental R-matrix the one of the 2-state six-vertex model.


\section{Integrable chiral Potts model}

\begin{figure}[htb]
\begin{center}
  \vspace*{0pt}
     \includegraphics[width=0.90\hsize]{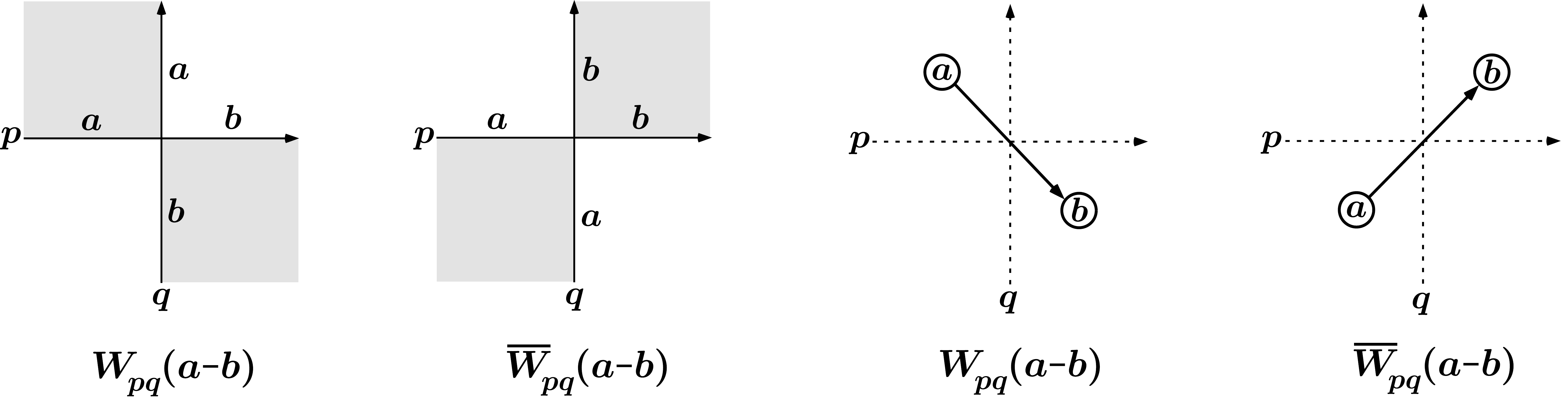}
\end{center}
\caption{Graphical representation of the two types of chiral Potts
Boltzmann weights with spin states $a,b=1,\ldots,N$ and oriented
rapidity lines $p,q$. Both the checkerboard vertex model (left)
and spin model (right) representations are given.}
\label{fig3}
\end{figure}

The $N$-state integrable chiral Potts model is defined by its
Boltzmann weights \cite{BPA},
\be
\displaystyle{W_{pq}(n)\over W_{pq}(0)}=
\prod_{j=1}^n\,{d_pb_q-a_pc_q\omega^j\over b_pd_q-c_pa_q\omega^j},
\quad
\displaystyle{\wb_{pq}(n)\over\wb_{pq}(0)}=
\prod_{j=1}^n\,{\omega a_pd_q-d_pa_q\omega^j\over c_pb_q-b_pc_q\omega^j},
\ee
see also Fig.~\ref{fig3}. Here the rapidities
$p=(a_p,b_p,c_p,d_p)$ and $q=(a_q,b_q,c_q,d_q)$
live on the chiral Potts curve:
\be\fl
a_p^{\,N}+k'b_p^{\,N}=k\,d_p^{\,N},\quad
k'a_p^{\,N}+b_p^{\,N}=k\,c_p^{\,N}, \quad
k^2+k'{}^2=1,\quad\omega\equiv\mathrm{e}^{2\pi\mathrm{i}/N}.
\ee
These Boltzmann weights satisfy the star-triangle equation
represented in Fig.~\ref{fig4}, see the appendix of \cite{AP}.

\begin{figure}[htb]
\begin{center}
  \vspace*{0pt}
     \includegraphics[width=0.45\hsize]{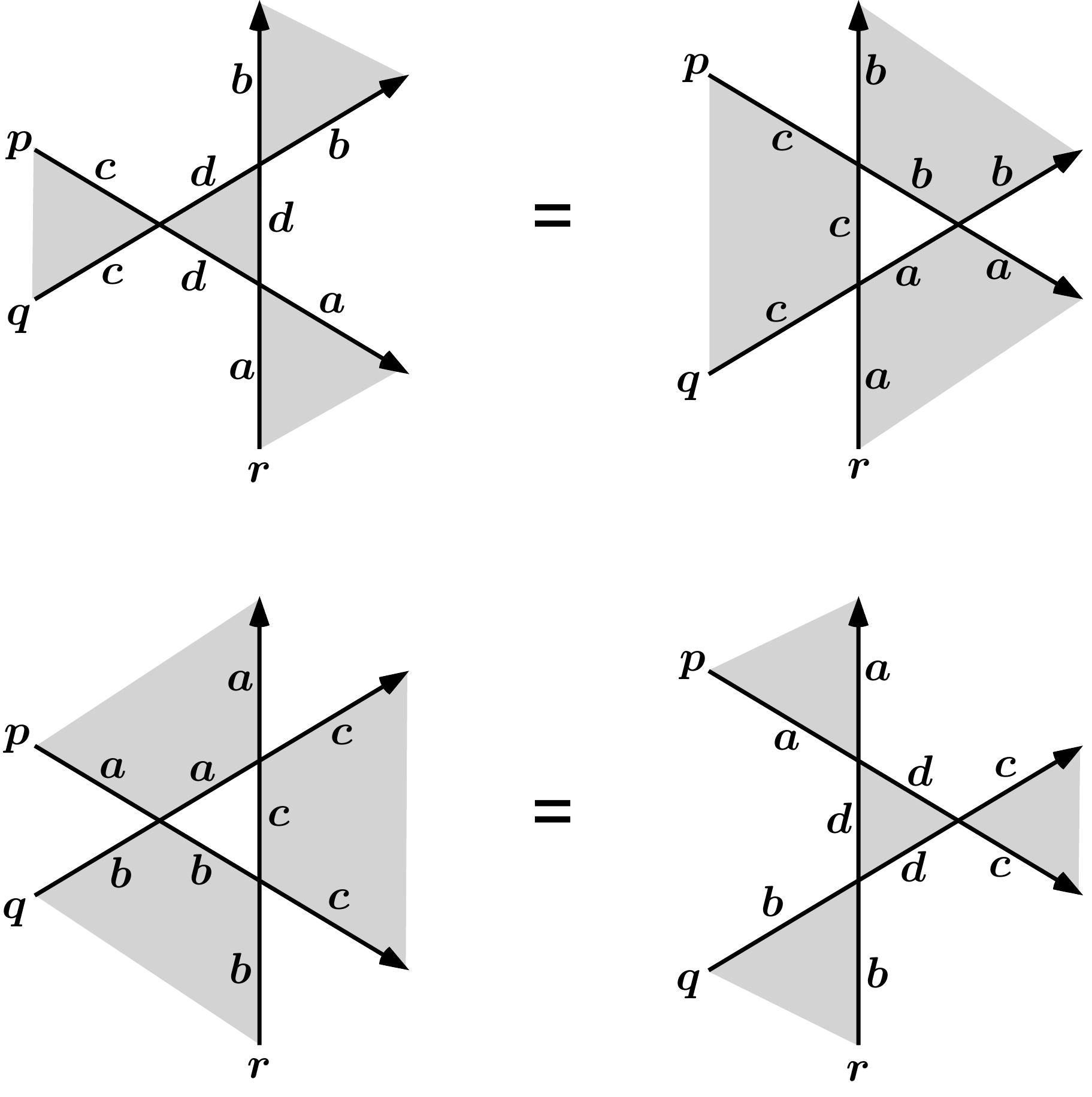}\qquad
        \includegraphics[width=0.45\hsize]{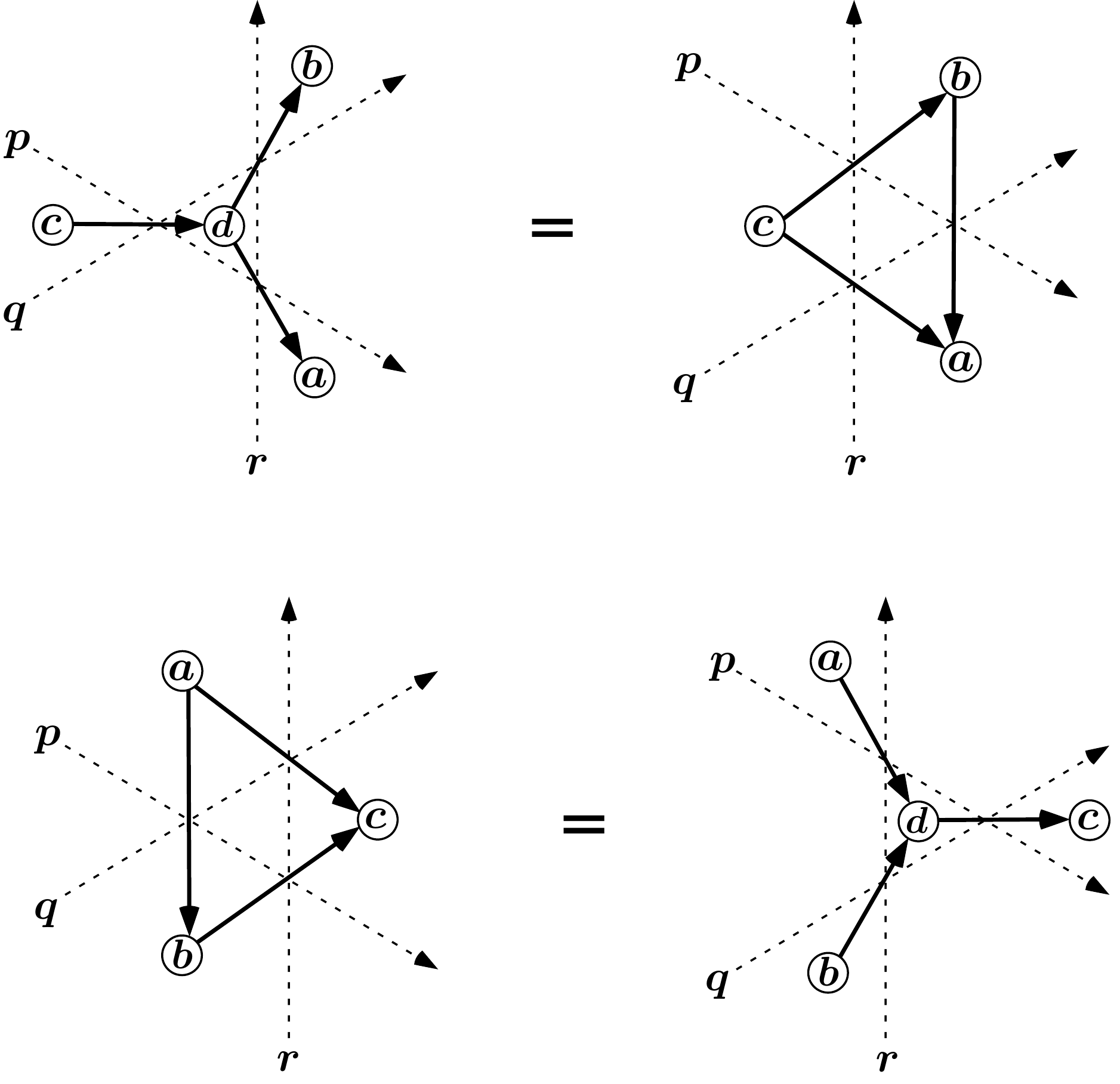}
\end{center}
\caption{Graphical representations of the checkerboard Yang--Baxter
equation on the left and the equivalent star-triangle equation on
the right.}
\label{fig4}
\end{figure}

\begin{figure}[htb]
\begin{center}
  \vspace*{0pt}
     \includegraphics[width=0.45\hsize]{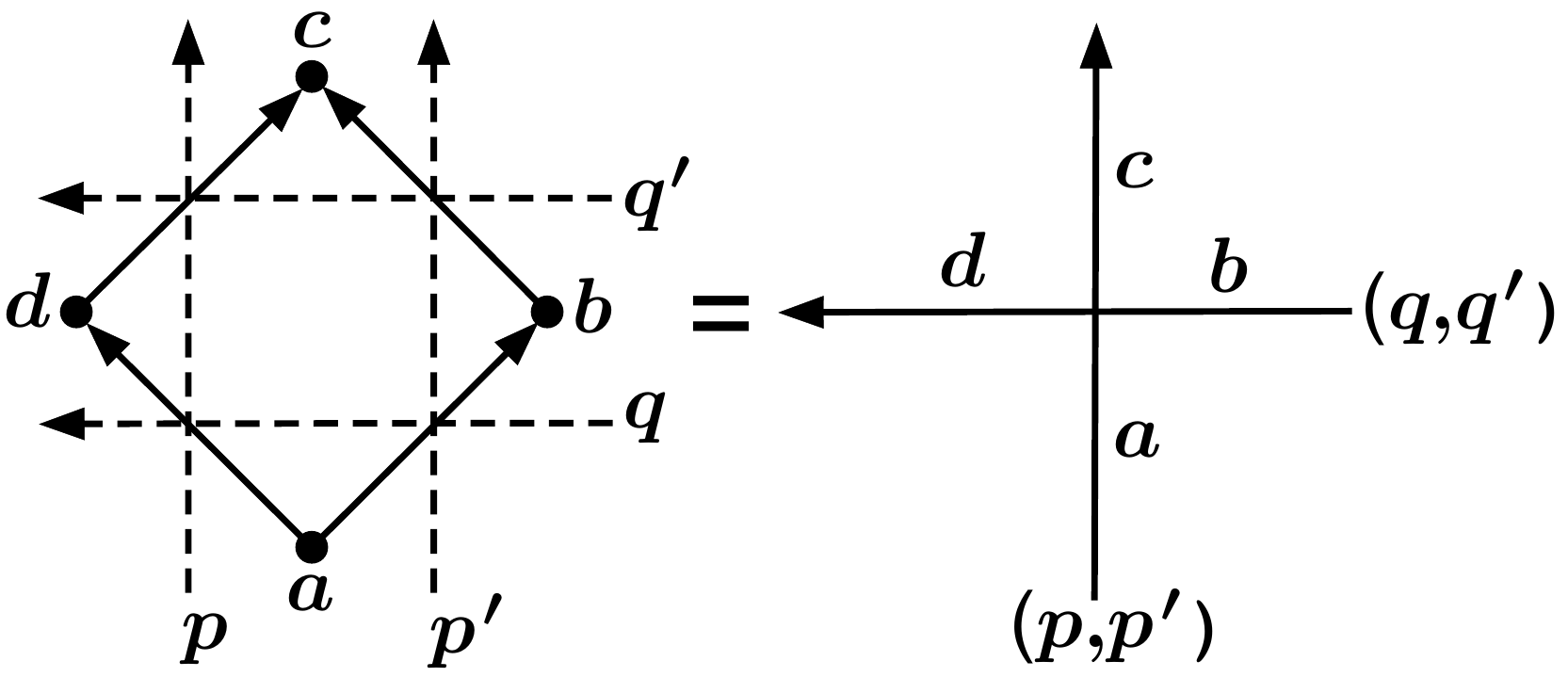}\qquad
        \includegraphics[width=0.45\hsize]{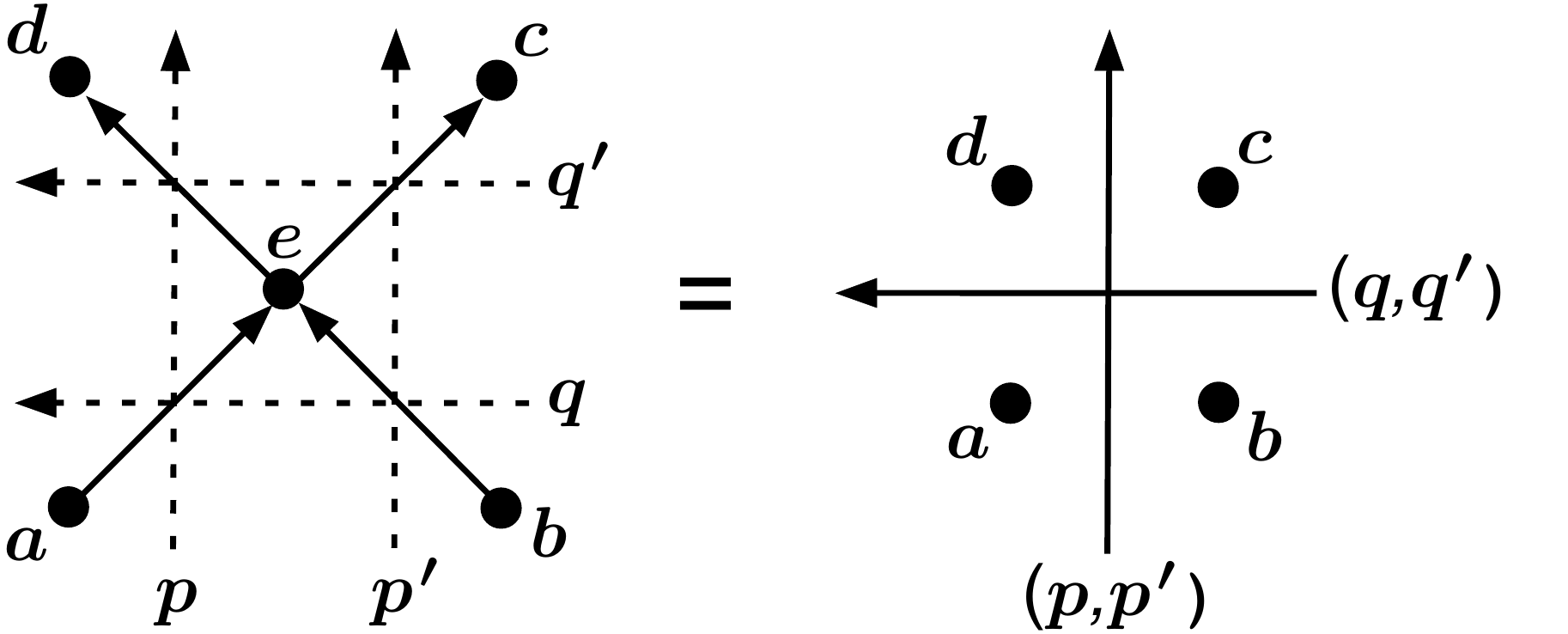}
\end{center}
\caption{The diamond (left) and the star (right) of four Boltzmann weights.}
\label{fig5}
\end{figure}

\begin{figure}[htb]
\begin{center}
  \vspace*{0pt}
     \includegraphics[width=0.75\hsize]{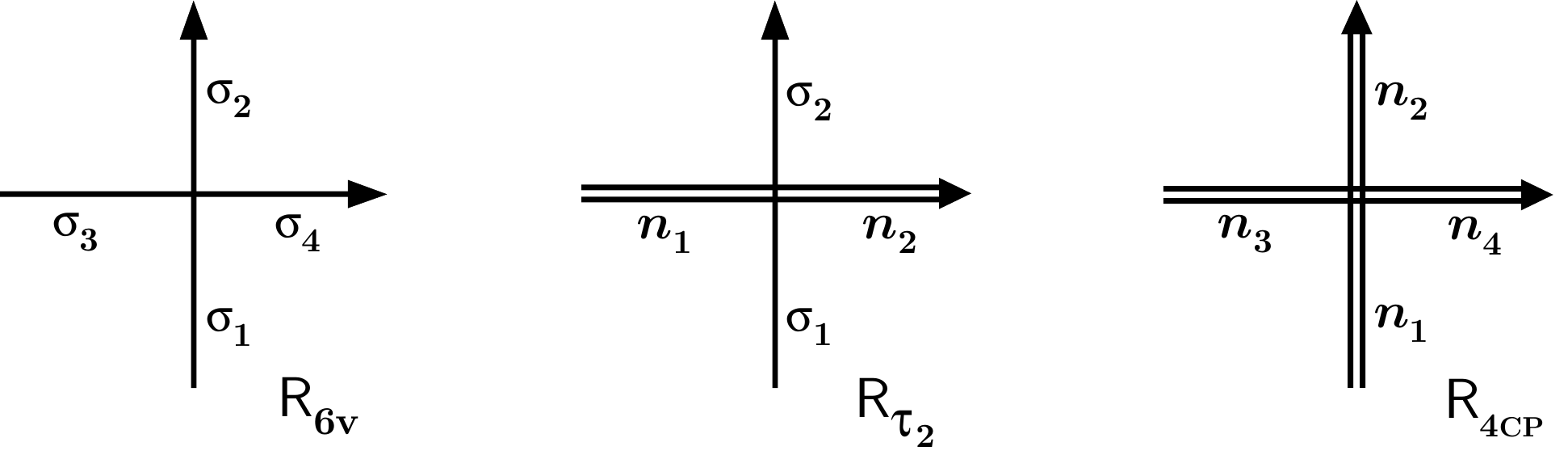}
\end{center}
\caption{The three kinds of R-matrices to be used.
Here all $\sigma_i=0,1$, correspond to the spin-$1\over2$ 
representation, whereas all $n_i=0,\cdots,N\!-\!1$,
i.e. $n_i\in\mathbb{Z}_N$, correspond to the cyclic representation.}
\label{fig6}
\end{figure}

Combining four chiral Potts Boltzmann weights as a diamond or
a star as in Fig.~\ref{fig5}, we get R-matrices satisfying the
uniform Yang--Baxter equation, so that we can forget about the
checkerboard shading. Bazhanov and Stroganov \cite{BS} used the diamond
map to relate chiral Potts with the six-vertex model for $N=\,$odd.
Baxter, Bazhanov and Perk \cite{BBP} used the star map instead to relate
chiral Potts with the six-vertex model for all $N$. Their resulting
interaction-round-a-face (IRF) model can be mapped to a vertex model,
see $\mathsf{R}_{\mathrm{4CP}}$ in Fig.~\ref{fig6},
using a Wu-Kadanoff-Wegner map \cite{Wu,KW}, putting now the differences
$n_1=a-b$, $n_2=d-c$, $n_3=a-d$, $n_4=b-c$ (mod $N$) on the four edges.

In quantum-group representation theory, the fundamental R-matrix
$\mathsf{R}_{\mathrm{6v}}$ intertwines two spin-$\frac12$
representations and $\mathsf{R}_{\mathrm{4CP}}$ intertwines two
(minimal) cyclic representations. We need one more R-matrix
$\mathsf{R}_{\tau_2}$ interwining the two different types of
representations, see Fig.~\ref{fig6}. This R-matrix generates
what is now often called a $\tau_2$ model, a name going back to
\cite{BS,BBP}, where a spin-$S$ representation intertwined with
a cyclic representation corresponds with a $\tau_{2S+1}$ transfer matrix.
The three R-matrices $\mathsf{R}_{\mathrm{6v}}$,
$\mathsf{R}_{\tau_2}$ and $\mathsf{R}_{\mathrm{4CP}}$ satisfy
a succession of four Yang--Baxter equations represented in
Fig.~\ref{fig6}. Here single rapidity lines correspond to
spin-$\frac12$ representations of
$\mathrm{U}_q({\widehat{\hbox{\bfrak sl}}}(2,\mathbb{C}))$, or
quantum affine SL(2). Double rapidity lines carry two chiral Potts
rapidities $(p,p')$ and correspond to a minimal cyclic representation
of $\mathrm{U}_q({\widehat{\hbox{\bfrak sl}}}(2,\mathbb{C}))$.
This requires $q$ to be a root of unity, say
$q=\omega=\mathrm{e}^{2\pi\mathrm{i}/N}$.

\begin{figure}[h]
\begin{center}
  \vspace*{0pt}
     \includegraphics[width=0.7\hsize]{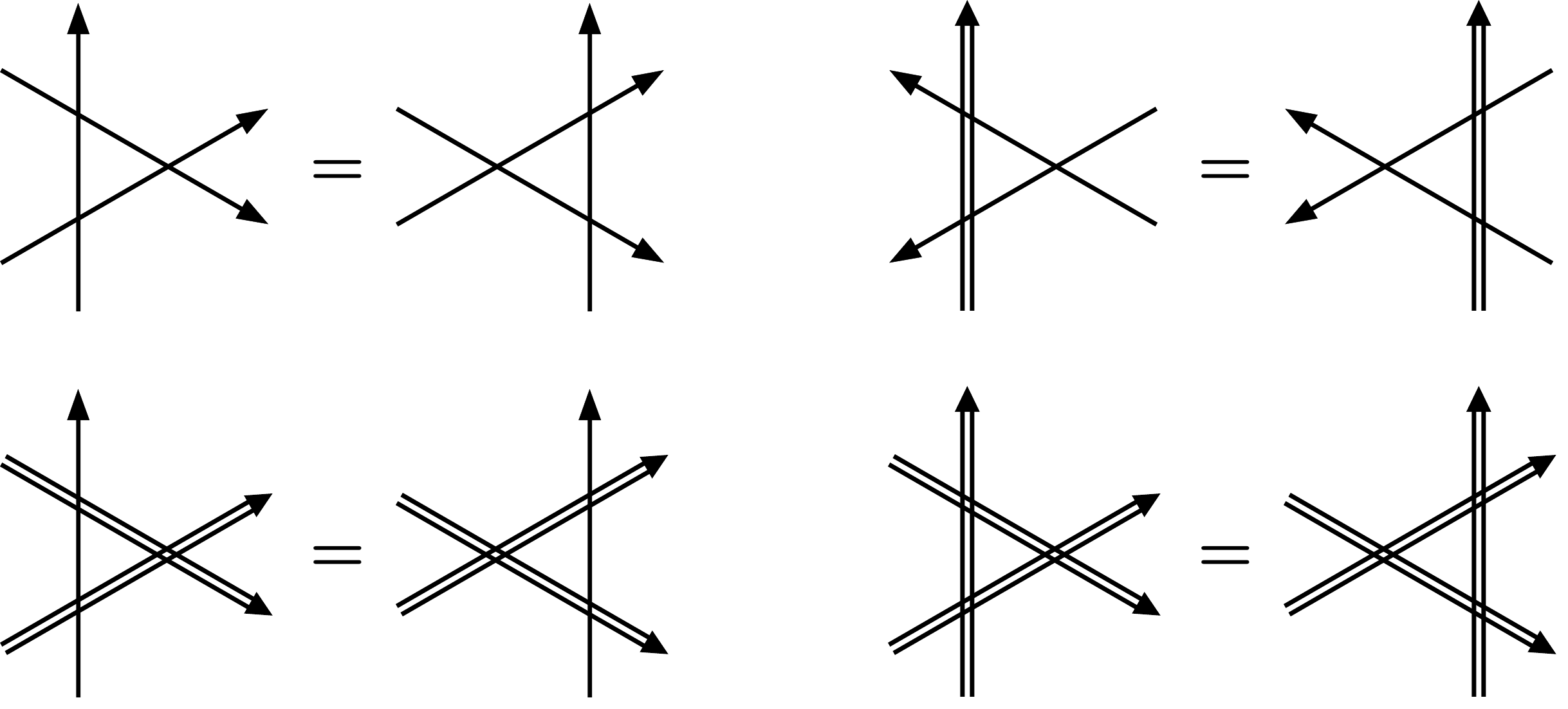}
\end{center}
\caption{The four different Yang--Baxter equations.}
\label{fig7}
\end{figure}


\section{The Boltzmann weights of the six-vertex model}

\begin{figure}[h]
\begin{center}
  \vspace*{0pt}
     \includegraphics[width=0.6\hsize]{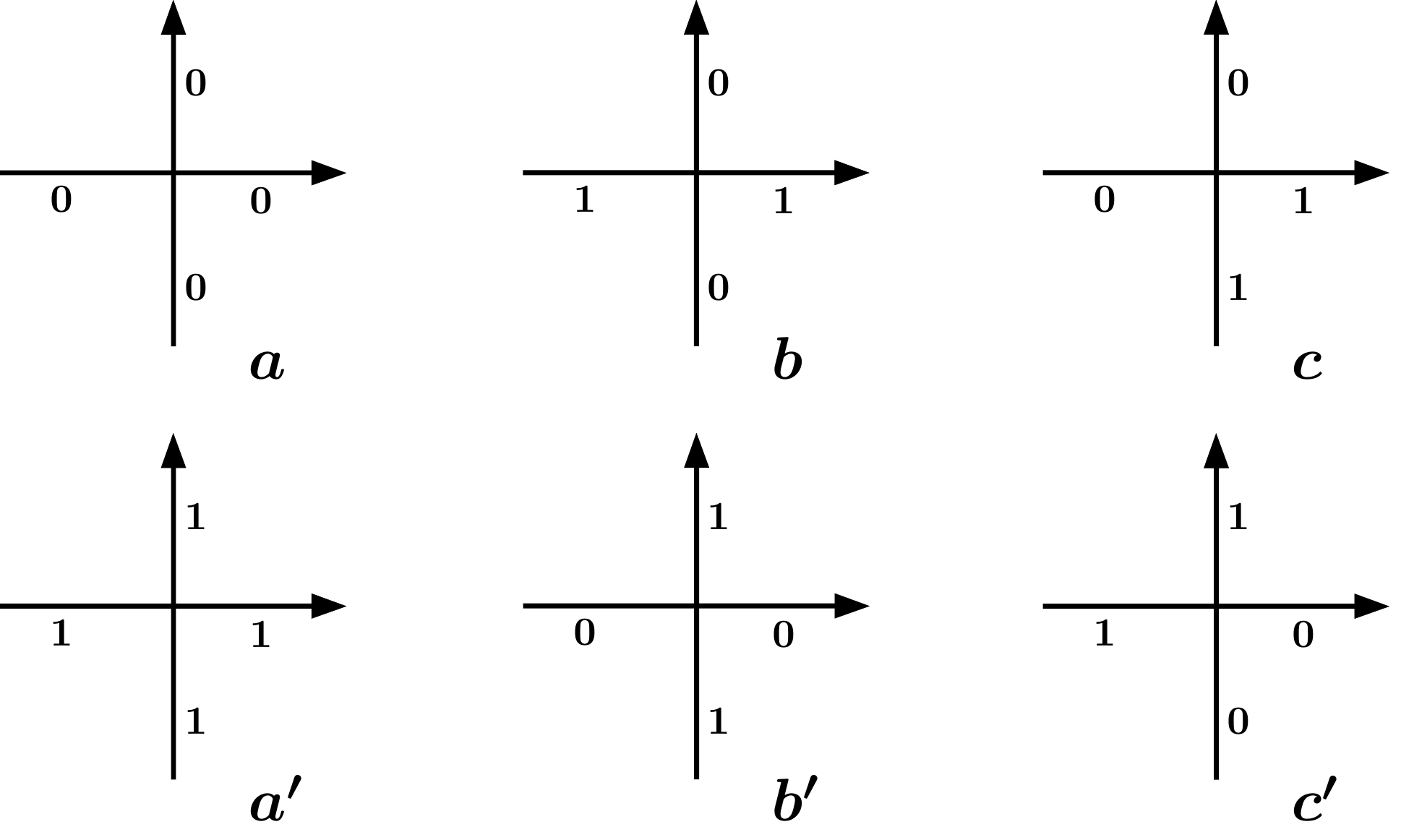}
\end{center}
\caption{The six different six-vertex Boltzmann weights.}
\label{fig8}
\end{figure}

The most general six-vertex model has six different weights as
given in Fig.~\ref{fig8}. This is the case $n=2$, $m=0$ of the previous
section and now we can absorb the twisting factor $G_{ab}$ in
(\ref{eq1b}) and the exponential factor
${\rm e}^{(p_0-q_0){\rm sign}(a-b)}$ in (\ref{eq1c}) into the
gauge rapidities $p_{\pm i}$ and $q_{\pm i}$. We also go to the
trigonometric representation replacing sinh by sin and we relabel
the states $a,b=1,2$ as $\sigma,\sigma'=0,1$. Different gauge
choices lead to different $\tau_2$ models that have been connected
with chiral Potts.

In the symmetric six-vertex model one has
$a'=a$, $b'=b$, $c'=c$. With this start Korepanov%
\footnote{See \cite{Korepanov,Korepanov2} and references cited
in \cite{Perk}.}\ found
a $\tau_2$ model, but no chiral Potts. To understand why, we
parametrize the weights of the symmetric six-vertex model as
\be
a=\mathcal{N}\sin(\eta+(v-u)),\quad
b=\mathcal{N}\sin(v-u),\quad
c=\mathcal{N}\sin(\eta),
\ee
with additive rapidities $u$ and $v$. There is also a multiplicative parametrization,
\be
q\equiv\mathrm{e}^{2\mathrm{i}\eta},\quad
x=\mathrm{e}^{2\mathrm{i}u},\quad y=\mathrm{e}^{2\mathrm{i}v},\quad
\mathcal{C}=\mathcal{N}\,{q^{1/2}\over\raise1pt\hbox{$2\mathrm{i}$}}
\Big(\frac{y}{x}\Big)^{\!1/2},
\ee
so that
\be
a=\mathcal{C}\,
\Big(1-q^{-1}\XY\Big),\quad
b=\mathcal{C}\,q^{-1/2}
\Big(1-\XY\Big),\quad
c=\mathcal{C}\,\big(1-q^{-1}\big)
\Big(\XY\Big)^{\!1/2}.
\ee
If one sets $\eta=j\pi/N$, then one finds
$q\equiv\mathrm{e}^{2\mathrm{i}\eta}=
\mathrm{e}^{2j\pi\mathrm{i}/N}$, the root-of-unity case,
which is one way to arrive at cyclic representations of quantum groups.
However, the symmetric gauge is not a good start for the fundamental
representation of sl(2) quantum: The square root $\sqrt{x/y}$ makes
things ugly and it could have been eliminated by a gauge transformation.
Up to normalization $\mathcal{C}$ the R-matrix used by Korepanov is
\be\fl
\mathsf{R}_{\mathrm{sym}}(x,y)=\pmatrix{1-\XY q^{-1}&0&0&0\cr
0&\Big(1-\XY\Big)\, q^{-1/2}&\Big(\XY\Big)^{\!1/2}(1-q^{-1})&0\cr
0&\Big(\XY\Big)^{\!1/2}(1-q^{-1})&\Big(1-\XY\Big)\, q^{-1/2}&0\cr
0&0&0&1-\XY q^{-1}}.
\ee
The $\big(x/y)^{1/2}$ and $q^{-1/2}$\ cause complications
especially for $N$ even.

Bazhanov and Stroganov \cite{BS} used the asymmetric gauge
typically used in quantum group theory,
\be\fl
\mathsf{R}_{\mathrm{B\& S}}(x,y)=\pmatrix{1-\XY q^{-1}&0&0&0\cr
0&\Big(1-\XY\Big)\, q^{-1/2}&\XY(1-q^{-1})&0\cr
0&1-q^{-1}&\Big(1-\XY\Big)\, q^{-1/2}&0\cr
0&0&0&1-\XY q^{-1}}.
\ee
They were able to arrive at the chiral Potts model only for $N$ odd.
Now the $q^{-1/2}$ still causes complications for $N$ even,
just like in the more general $n\geqslant2$ case of section 2.

A more asymmetric gauge was found in \cite{BBP} starting from the
chiral Potts side,
\be\fl
\mathsf{R}_{\mathrm{BBP}}(x,y)=\pmatrix{1-\XY q^{-1}&0&0&0\cr
0&1-\XY&\XY(1-q^{-1})&0\cr
0&1-q^{-1}&\Big(1-\XY\Big)q^{-1}&0\cr
0&0&0&1-\XY q^{-1}}.
\ee
This was already pointed out in \cite{Perk}. As now only
1, $x/y$, $q^{-1}$, and $(x/y)q^{-1}$ show up, the situation
is least complicated with the ``smallest linear dimension."
The commutation relations of the four elements of the monodromy
matrix are now least complicated \cite{Perk}.


\section{Gauge Changes of Six-Vertex Boltzmann Weights}

\begin{figure}[h]
\begin{center}
  \vspace*{0pt}
     \includegraphics[width=0.75\hsize]{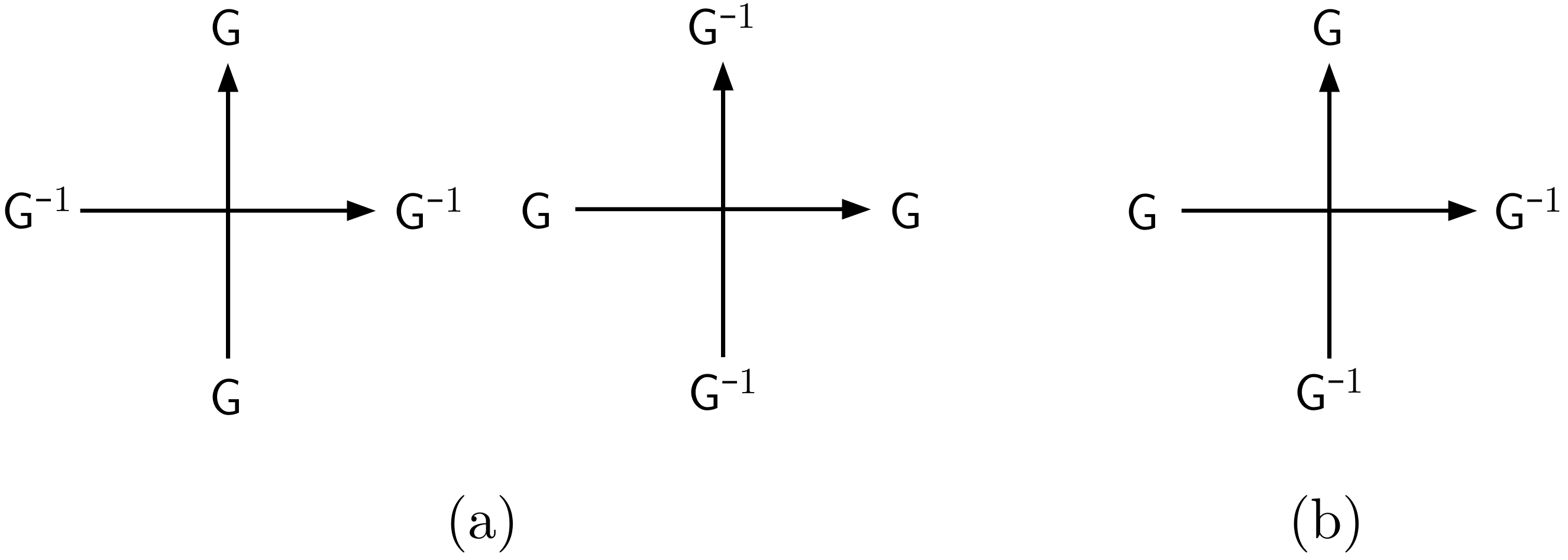} 
\end{center}
\caption{(a) Staggered gauge transform. (b) Uniform gauge transform.}
\label{fig9}
\end{figure}

In order to understand how the three approaches relate, we start
with $\mathsf{R}_{\mathrm{BBP}}$ and apply suitable gauge transforms
of the two types in Fig.~\ref{fig9}.
A staggered gauge transform with $\mathsf{G}$ of the simple
diagonal form
\be
\mathsf{G}=\pmatrix{\lambda&0\cr0&\lambda^{-1}},\quad \mbox{with}\quad \lambda=q^{1/8},
\ee
can be used to connect $\mathsf{R}_{\mathrm{B\&S}}$ and
$\mathsf{R}_{\mathrm{BBP}}$\strut\ in each of two different ways
given in Fig.~\ref{fig9}(a).
\medskip\noindent
A uniform gauge transform
\be
\mathsf{G}=\pmatrix{\lambda&0\cr0&\lambda^{-1}},\quad \mbox{with}\quad \lambda=\Big(\frac{x}{y}\Big)^{1/8},
\ee
as in Fig.~\ref{fig9}(b) connects
$\mathsf{R}_{\mathrm{sym}}$ and $\mathsf{R}_{\mathrm{B\&S}}$.
\medskip

In the approach of BBP \cite{BBP} there is no difficulty with even
roots of unity. However, the staggered gauge transforms to the
Bazhanov--Stroganov approach, and then also to the Korepanov
symmetric gauge, lead to complications: Two distinct $\tau_2$
matrices arise in the
$\mathsf{R}_{\mathrm{6v}}\mathsf{R}_{\tau_2}\mathsf{R}_{\tau_2}$
Yang--Baxter equation of Fig~\ref{fig7}.

It may be said that Korepanov \cite{Korepanov,Korepanov2}
during 1986--1987 has made some start to solve the even root-of-unity problem using two $\tau_2$ matrices. He solved the
$\mathsf{R}_{\mathrm{6v}}\mathsf{R}_{mathrm{6v}}\mathsf{R}_{\tau_2}$
Yang--Baxter equation of Fig~\ref{fig7} using
$\mathsf{R}_{\mathrm{sym}}$, giving one $\mathsf{R}_{\tau_2}$
for $N=\,$odd, while for $N=\,$ even his solution has
two different $\mathsf{R}_{\tau_2}$. However, he did not address
the next steps in Fig~\ref{fig7}, so that he could not arrive at
the chiral Potts model.

Bazhanov and Stroganov did address the next steps in the succession
of Yang--Baxter equations, starting with $\mathsf{R}_{\mathrm{B\&S}}$,
which is the typical choice for the intertwiner of two fundamental
representations of
$\mathrm{U}_q({\widehat{\hbox{\bfrak sl}}}(2,\mathbb{C}))$.
However, to explicitly represent $\mathsf{R}_{\tau_2}$ for
$q=\omega\equiv\mathrm{e}^{2\pi\mathrm{i}/N}$,
they introduce $q_1\vp=q^{(N+1)/2}$, satisfying $q_1^N=1$, $q=q_1^{-2}$,
which can only be done for $N=\,$odd. For $N=\,$even, both solutions
$q_1\vp$ of $q_1^{\,2}=q$ satisfy $q_1^N=-1$.

Finally, the two approaches of B\&S and BBP lead to different $q$-Pochhammer symbols, as explained in \cite{Perk}, namely
\be
[a;q_1]_n=\prod_{k=1}^n(a^{-1}q_1^{n-1}-aq_1^{1-n})
\quad\hbox{versus}\quad
(a;q)_n=\prod_{k=1}^n(1-aq^{n-1}),
\ee
and also different $q$-integers,
\be
\displaystyle[q_1]_n={q_1^n-q_1^{-n}\over q_1\vp-q_1^{-1}}
\quad\hbox{versus}\quad
\displaystyle(q)_n={1-q^n\over1-q}.
\ee
The approach of \cite{BBP} leads to more standard notations
of the theory of basic hypergeometric functions.


\end{document}